\documentstyle[12pt]{article}
\def\hybrid{\topmargin -20pt	\oddsidemargin 0pt
	\headheight 0pt	\headsep 0pt
	\textwidth 6.25in	% A4 paper
	\textheight 9.5in	% A4 paper
	\marginparwidth .875in
	\parskip 5pt plus 1pt	\jot = 1.5ex}

\hybrid
\newskip\humongous \humongous=0pt plus 1000pt minus 1000pt

\newif\ifdtup

\relax

\def\e{\epsilon}
\def\de{\delta}
\def\dd{\bar\delta}
\def\be{\begin{equation}}
\def\ee{\end{equation}}
\def\ba{\begin{eqnarray}}
\def\ea{\end{eqnarray}}
\def\d{\partial}
\def\db{\bar{\partial}}

\def\a{\alpha}

\def\s{\sigma}

\def\e{\epsilon}

\def\rr{{\rm regular}}
\def\xx{\hbox{ }^*_*}
\def\ec{\hat E^{c}_{2}}

\begin{document}
\renewcommand{\theequation}{\thesection.\arabic{equation}}
\newcommand{\beq}{\begin{equation}}
\newcommand{\eeq}[1]{\label{#1}\end{equation}}
\newcommand{\ber}{\begin{eqnarray}}
\newcommand{\eer}[1]{\label{#1}\end{eqnarray}}
\begin{titlepage}
\begin{center}

\hfill CERN-TH.7059/93\\
\hfill LPTENS-93/42\\
\hfill hep-th/9310202\\

\vskip .9in

{\large \bf  String Propagation in Gravitational Wave Backgrounds
}
\vskip .9in

\vskip .15in

{\bf Elias Kiritsis and Costas Kounnas} \footnote{e-mail addresses:
KIRITSIS@NXTH04.CERN.CH, KOUNNAS@NXTH04.CERN.CH}\\
\vskip
 .1in

{\em Theory Division, CERN, CH-1211\\
Geneva 23, SWITZERLAND}\\

\vskip .5in

\end{center}

\vskip .4in

\begin{center} {\bf ABSTRACT } \end{center}
\begin{quotation}\noindent
The Conformal Field Theory of the current algebra of the centrally
extended 2-d Euclidean group
is analyzed. Its representations can be written in terms of four free
fields (without background charge) with signature ($-$+++).
We construct all irreducible representations of the current algebra
with unitary base out of the free fields and their orbifolds.
This is used to investigate the spectrum and scattering of strings
moving in the background of a gravitational wave.
We find that all the dynamics happens in the transverse space or the
longitunal one but not both.

\end{quotation}
\vskip 3.0cm
CERN-TH.7059/93 \\
October 1993\\
\end{titlepage}
\vfill
\eject
\def\baselinestretch{1.2}
\baselineskip 16 pt
\setcounter{equation}{0}

Current algebra \cite{c1,c2} has proved to be  a valuable tool in
understanding
Conformal Field Theory (CFT) and String Theory.
All the conformal field theories we know so far can be constructed
one way or another from current algebras.
The simplest are the WZW models \cite{W1,KZ}, which realize current
algebra as their full symmetry.
Gauging a subgroup, we obtain the coset models
\cite{c2,coset1,coset2}.
However, more general constructions of CFTs can be effected using
current
algebra \cite{HK}, which provide examples of truly irrational CFTs
\cite{HK2}.
In such constructions the key ingredient is the factorization of a
CFT,
using its (2,0) operators \cite{K1}.

For obvious reasons, the first type of algebras to be analysed were
compact ones, used for compactification purposes in String Theory
\footnote{They can also play a role, though, as parts of the
non-compact spacetime.}.
Later on, non-compact algebras (of the type SL(N,R), SU(M,N) and
SO(M,N)) and their cosets have been considered
\cite{nc} in order to describe curved Minkowski signature spacetimes.
Only recently did current algebras of the non-semisimple type receive
some
attention \cite{D,W2}.
We will analyse here the CFT description of such algebras and their
cosets,
which will reveal many differences from current algebras previously
considered. Their spacetime interpretation is special, and it turns
out, in the simplest models, that it is associated with the existence
of null Killing vectors \cite{W2}.

Below, we will consider the current algebra theory in \cite{W2},
which corresponds
to a central extension of the 2-d Euclidean group $E_{2}$, which we
denote
by $E^{c}_{2}$.
We will show that the algebra is described in terms of four free
fields without
background charge, with signature ($-$+++).
We will construct its representations and use our results towards
understanding the spectrum and scattering of (bosonic) strings moving
in a gravitational wave background.It turns out that non-trivial
interactions occur either in  the transverse space or in the
logitudinal directions but not both.
We will also examine cosets of the $\ec$ theory.

We start with the current algebra $\hat E_{2}$.
This is specified by the OPEs
$$J_{a}(z)J_{b}(w)={G_{ab}\over (z-w)^2}+{f_{ab}}^{c}{J_{c}(w)\over
(z-w)}+
\rr,\eqno(1)$$
where $(J_{1},J_{2},J_{3})\sim(P_{1},P_{2},J)$, the $P_{i}$
generating the translations and $J$ being the generator of rotations.
The only non-zero structure constants are ${f_{31}}^{2}=1$ and
${f_{32}}^{1}=-1$;
$G_{ab}$ is an invariant bilinear form (metric) of the algebra. $G$
is a symmetric matrix
and the Jacobi identities
constrain it so that $f_{abc}\equiv {f_{ab}}^{d}G_{dc}$ is completely
antisymmetric.
The only such invariant bilinear form for $E_{2}$ is
$$G=\left(\matrix{0&0&0\cr 0&0&0\cr 0&0&k\cr}\right).\eqno(2)$$
However, this bilinear form is degenerate, and this is responsible
for the peculiar properties of $\hat E_{2}$, as we will see.

When we start from a current algebra, we can construct stress tensors
that are bilinear in the currents\footnote{Conventions on normal
ordering can be found in \cite{HK}.},
$T_{L}=L^{ab}\xx J_{a}J_{b}\xx$.
The coefficients $L^{ab}$ must satisfy the affine-Virasoro Master
Equation
(ME) \cite{HK}:
$$L^{ab}=2L^{ac}G_{cd}L^{db}+L^{cd}L^{ef}{f_{ce}}^{a}{f_{df}}^{b}+
\left(L^{cd}{f_{ce}}^{f}{f_{df}}^{a}L^{be}+(a\leftrightarrow
b)\right).\eqno(3)$$
For any solution $L_{*}$ of the ME, the central charge of $T_{L_{*}}$
is
$$c(L_{*})=2G_{ab}L_{*}^{ab}\;\;.\eqno(4)$$

For semisimple algebras, there is a (unique) solution to the ME, the
affine-Sugawara
(AS) construction, $L_{AS}^{ab}$, which has the following properties:

a.  All the currents $J_{a}$ are primary fields with respect to
$T^{AS}$
with conformal weight 1.

b.  For any solution $L_{*}$ to the ME, $L_{AS}-L_{*}$ is
automatically
another solution, and the two central charges add up to the
affine-Sugawara
central charge.

Property (a) is an important characteristic of the AS construction,
since
it implies that the current algebra symmetry remains unmolested in
the quantum
theory.

It is not difficult to discern for the case of $\hat E_{2}$ that
there is no
analogue of the AS construction that satisfies property (a) above.
The best we can do is to find a stress tensor with respect to which
$J$
is primary with $\Delta =1$ and $P_{i}$ having $\Delta=0$.
This is no surprise since the WZW model associated with $E_{2}$ is
1-dimensional due to the degeneracy of the bilinear form.

The way out is to add a central element, $J_{4}=T$, in the Lie
algebra
of $E_{2}$, as was done in \cite{W2}, to obtain the centrally
extended Lie algebra
$E^{c}_{2}$.
Thus, the commutation relation $[P_{1},P_{2}]=0$ is modified to
$[P_{1},P_{2}]=T$, and $T$ commutes with everything.
This is the Euclidean algebra in 2-d for a charged particle in a
constant
magnetic field.
The most general invariant bilinear form for $E^{c}_{2}$ is
$$G=k\left(\matrix{1&0&0&0\cr 0&1&0&0\cr 0&0&b&1\cr
0&0&1&0\cr}\right).\eqno(5)$$

For $\hat E^{c}_{2}$ there is a unique solution to the ME, which has
all the
properties (a+b) of the AS construction. It is given by:
$$L_{AS}={1\over 2k}\left(\matrix{1&0&0&0\cr 0&1&0&0\cr 0&0&0&1\cr
0&0&1&-b+{1\over k}\cr}\right)={1\over 2}G^{-1}+{1\over 2k^2}
\left(\matrix{0&0&0&0\cr 0&0&0&0\cr 0&0&0&0\cr
0&0&0&1\cr}\right),\eqno(6)$$
with $c=4$.
We see that although the stress tensor has an extra one-loop
contribution
to its classical value, ${1\over 2}G^{-1}$, the value of the central
charge is entirely classical.

We will proceed to analyze further the structure of $\ec$.
We start by bosonizing the Cartan subalgebra, spanned by $J$ and $T$.
Defining
$$J^{+}={1\over \sqrt{kb}}J\;\;\;,\;\;\;
J^{-}={1\over \sqrt{kb}}\left(J-bT\right)\eqno(7)$$
we have diagonalized the Cartan subalgebra,
$$J^{+}(z)J^{+}(w)=-J^{-}(z)J^{-}(w)={1\over (z-w)^2}+\rr\;\;,\;\;
J^{+}(z)J^{-}(w)=\rr.\eqno(8)$$
Thus, we can express $J^{\pm}$ in terms of two bosonic fields,
$x^{0},x^{3}$ as
$$J^{+}(z)=\d x^{0}\;\;\,\;\;J^{-}(z)=\d x^{3},\eqno(9)$$
where the bosons have free 2-point functions\footnote{We focus only
on the holomorphic part from now on.}
$$\langle x^{0}(z)x^{0}(w)\rangle =-\langle x^{3}(z)x^{3}(w)\rangle
=\log(z-w)\;\;,\;\;\langle x^{0}(z)x^{3}(w)\rangle =0.\eqno(10)$$
The AS stress tensor (6) becomes
$$T_{AS}={1\over 2k}\xx P_{1}^2+P_{2}^2\xx+{1\over 2}\d x^{+}\d
x^{-}+
{1\over 2Q^2}(\d x^{-})^2,\eqno(11)$$
where we defined light-cone coordinates $x^{\pm}=x^{0}\pm x^{3}$ and
$Q=\sqrt{kb}$.
We have assumed without loss of generality that $k>0,b>0$.
For $b<0$ the roles of $x^{0}$ and $x^{3}$ are interchanged.

The rest of the algebra consists of a ``raising" generator
$P^{+}=P_{1}+iP_{2}$
and its corresponding ``lowering" operator $P^{-}=P_{1}-iP_{2}$.
They are charged under the Cartan action
$$J^{+}(z)P^{\pm}(w)={\mp i\over Q}{P^{\pm}(w)\over (z-w)}+\rr
\;\;,\;\;
J^{-}(z)P^{\pm}(w)={\mp i\over Q}{P^{\pm}(w)\over
(z-w)}+\rr.\eqno(12)$$
We see that the Lie algebra $E^{c}_{2}$ has a single root of length
square zero.

The next step is to do the analogue of the parafermionic
decomposition
of standard current algebra \cite{fz,g}.
We write
$$P^{\pm}(z)=\exp\left[{\mp i\over
Q}x^{-}\right]V^{\pm}(z).\eqno(13)$$
Then, the OPEs (12) imply that $V^{\pm}$ do not depend on $x^{\pm}$.
Finally, the $P^{\pm}$ OPEs
$$P^{+}(z)P^{-}(w)={2k\over (z-w)^2}-2i{k\over Q}{\d x^{-}\over
(z-w)}+\rr
\eqno(14a)$$
$$P^{+}(z)P^{+}(w)=\rr\;\;\;,\;\;\;P^{-}(z)P^{-}(w)=\rr\eqno(14b)$$
imply that
$$V^{+}(z)V^{-}(w)={2k\over (z-w)^2}+2kT_{2}(w)+{\cal
O}(z-w)\eqno(15a)$$
$$V^{+}(z)V^{+}(w)=\rr\;\;\;,\;\;\;V^{-}(z)V^{-}(w)=\rr,\eqno(15b)$$
where we have explicitly indicated the ${\cal O}(1)$ contribution
in the $V^{+}V^{-}$ OPE for future convenience.
A straightforward computation gives
$${1\over 2k}\xx P^{2}_{1}+P_{2}^{2}\xx =T_{2}-{1\over 2Q^2}(\d
x^{-})^2.\eqno(16)$$
Putting together eqs. (11) and (16), we observe that the AS stress
tensor has factorized:
$$T_{AS}=T_{2}+{1\over 2}\d x^{+}\d x^{-}.\eqno(17)$$
Thus, $T_{2}$ satisfies the Virasoro algebra with $c=2$.
This, and the OPEs (15a,b) imply (uniquely) that $T_{2}$ and
$V^{\pm}$ can be bosonized in terms of two more positive
signature\footnote{A signature (+$-$) is also possible here. However,
we do not choose it because we would like the full theory to have
Minkowski signature.}free bosons, $x^{1}$ and $x^{2}$, as
$$V^{\pm}=i\sqrt{k}\left(\d x^{1}\pm i\d x^{2}\right)\;\;,\;\;
T_{2}=-{1\over 2}(\d x^{1})^2-{1\over 2}(\d x^{2})^2\eqno(18)$$
with
$$\langle x^{i}(z)x^{j}(w)\rangle =-\delta^{ij}\log(z-w)\;\;\;,\;\;\;
i,j=1,2.\eqno(19)$$
The AS stress tensor is thus written entirely in terms of four free
bosons
of signature ($-$+++).

We note that only the variable $Q$ has remained as a parameter in the
current algebra.
If, $Q=0$, then the current algebra and its representations are that of
flat 4-d Minkowski space.
If $Q>0$, for non-compact $x^{\pm}$, it can always be set to one with a

Lorentz boost.

Having the free field representation of the current algebra $\ec$, we
can
construct representations.
We will first describe the unitary irreducible representations of the
$E^{c}_{2}$ Lie
algebra\footnote{This algebra is also known as the ``oscilator"
algebra
since it is the algebra of the phase space operators $P\sim
P_{1},Q\sim P_{2}$, the harmonic oscilator Hamiltonian $H\sim J$ and
the central element $i\hbar \sim T$.}. Such representations of the
zero mode algebra will (as usual)
serve as the base for the current algebra representations. They have
to be unitary otherwise there is no hope that negative norm states
will decouple in the string theory. The Virasoro constraints can
remove current algebra descendants but not primaries\footnote{Of
course the current algebra representations will not be unitary. This
non-positivity however can, in principle, be taken care of by the
Virasoro constraints.}.

The algebra $E^{c}_{2}$ has two independent Casimir operators. One is
linear
and is no other than the generator $T$. The other is quadratic and is
given
by
$$Q_{2}\equiv {1\over 2}(P^{+}P^{-}+P^{-}P^{+})+2JT,\eqno(20)$$
where, as usual, $P^{\pm}\equiv P_{1}\pm iP_{2}$.
Consider an eigenstate of the Cartan, $J,T$:
$$J|j,t> =ij|j,t>\;\;\;,\;\;\;T|j,t>=it|j,t>.\eqno(21)$$
The hermiticity property we have to choose for the raising operators
is $(P^{+})^{\dagger}=P^{-}$ while $J,T$ are anti-hermitian.
Then,
$$P^{\pm}|j,t>=\sqrt{Q_{2}+t(2j\mp 1)}|j\mp 1,t>.\eqno(22)$$
The only representation which has both a lowest and a highest weight
is the unit-like
representations: $Q_{2}=t=0$. They are characterized only by the
value of $j\in R$. They will be treated from now on, as special cases
of type I representations.

We have also unitary representations which have neither a highest nor
a lowest
weight (type I).
These are characterized by $t=0$, $Q_{2}>0$ and the fractional part
of $j$ since the spectrum of $iJ$ is integer spaced.

Unitary highest weight (hw) representations (type II).
For these, $Q_{2}=-t(2j-1)$ and they are characterized by $t>0$ and
$j\in C$.
The spectrum of $-iJ$ is $j,j-1,j-2,\cdots$.

Finally there are unitary lowest weight (lw) representations (type
III).
They have $Q_{2}=-t(2j+1)$ and they are characterized by $t<0$ and
$j\in C$.
The spectrum of $-iJ$  is $j,j+1,j+2,\cdots$\footnote{In string theory
$j$ should be real in order to have real conformal weights.}.

The above exhaust all the unitary irreducible representations of the
$E^{c}_{2}$ algebra\footnote{As the reader might have noticed, the
unitary representations of $E^{c}_{2}$ have a remarkable similarity
with SL(2,R)
representations. They are characterized though by an extra
parameter.}.
It is useful to calculate the value of the zero mode of the AS stress
tensor
given by
$${\cal{L}}_{0}^{AS}={1\over 2k}Q_{2}+{1\over 2k}\left({1\over
k}-b\right)T^{2}.\eqno(23)$$
$${\rm Type\;\;\;I}\;\;\;\;:\;\;\;\;\;{\cal{L}}_{0}^{AS}={1\over
2k}Q_{2}\phantom{aaaaaaaaaaaaaaaaaaaaaaaaa}\eqno(24a)$$
$${\rm Type\;\;\;II}\;\;:\;\;\;\;{\cal{L}}_{0}^{AS}=-{t\over
2k}\left(2j-1
+{t\over k}(1-bk)\right)\;\;,\;\;t>0\eqno(24b)$$
$${\rm Type\;\;\;III}\;:\;\;\;{\cal{L}}_{0}^{AS}=-{t\over
2k}\left(2j+1+{t\over k}(1-bk)\right)\;\;,\;\;t<0\eqno(24c)$$
\setcounter{footnote}{0}

Knowing the representations of $E^{c}_{2}$ we can use them as a base
in order
to construct the current algebra representation acting with the
negative modes of the currents.
In the current algebra Hilbert space they are annihilated by the
positive modes
of the curents.
This translates in the following OPE for the operators $R_{i}(z)$
that create out of the $E^{c}_{2}$ invariant vacuum the base states:
$$J_{a}(z)R_{i}(w)=T^{a}_{ij}{R_{j}(w)\over (z-w)}+\rr,\eqno(25)$$
where $T^{a}_{ij}$ are appropriate representation matrices for the
Lie algebra $E^{c}_{2}$.
We can immediately take care of the Cartan action (uniquely) by
taking
$R_{i}(z)$ to be of the form\footnote{Oscilators of $x^{\pm}$ would
always
be descedants of the Cartan algebra.}
$$R(z)\sim
\exp\left[ip_{+}x^{-}+ip_{-}x^{+}\right]V(x^{1},x^{2}).\eqno(26)$$
In this parametrization:
$$j=Q(p_{+}+p_{-})\;\;\;,\;\;\;t=2{k\over Q}p_{-}.\eqno(27)$$
We also have
$$P^{+}(z)\exp\left[ip_{+}x^{-}+ip_{-}x^{+}\right](w)\sim
(z-w)^{2p_{-}\over Q}\exp\left[i(p_{+}-1/Q)x^{-}
+ip_{-}x^{+}\right](w)+\cdots\eqno(28a)$$
$$P^{-}(z)\exp\left[ip_{+}x^{-}+ip_{-}x^{+}\right](w)\sim
(z-w)^{-{2p_{-}\over Q}}\exp\left[i(p_{+}+1/Q)x^{-}
+ip_{-}x^{+}\right](w)+\cdots\eqno(28b)$$

\underline{Type I Representations}.
Since for these representations $t=0$ this implies via (27) that
$p_{-}=0$.
At the base, they are generated by a single operator
$$V_{p_{+},\vec p_{T}}=\exp\left[ip_{+}x^{-}+i\vec p_{T}\cdot \vec
x_{T}\right].\eqno(29)$$
The full set of primary operators in this representation, generated
by the action of the zero-mode raising and lowering operators are
$$V_{n,p_{+},\vec p_{T}}=\exp\left[i(p_{+}+n/Q)x^{-}+i\vec p_{T}\cdot
\vec x_{T}\right]\eqno(30)$$
with $n\in Z$ and $0\leq p_{+}<1/Q$.
For $\vec p_{T}=0$ only the $n=0$ operator exists.
Their conformal weight under the AS stress tensor is $\Delta_{I}=\vec
p_{T}^{2}/2$ in agreement with (24a).
The multiplicities for fixed $J$ and $p_{+}\not=n/Q$ are given by
$1/\eta^{4}$.
The signature character (takes care also of the sign of norms of
states) is
$1/\eta^2$.

\underline{Type II Representations}. These have a hw state at the
base and we will write again  the hw operator as
$$R^{II}_{p_{+},p_{-}}=\exp\left[ip_{+}x^{-}+ip_{-}x^{+}\right]
V^{II}_{p_{-}}(x^{1},x^{2}).\eqno(31)$$
Here we have $p_{-}>0$.
The OPE (25) implies that
$$\d x_{T} (z) V^{II}_{p_{-}}(w)\sim (z-w)^{-{2p_{-}\over
Q}}\;\;,\eqno(32a)$$
$$\d \bar x_{T} V^{II}_{p_{-}}(w)\sim (z-w)^{-1+{2p_{-}\over
Q}},\eqno(32b)$$
where $x_{T}=x^{1}+ix^{2}$.
It is obvious from eqs. (32a,b) that the operator $V^{II}_{p_{-}}$ is
an operator that twists the transverse coordinates via the
transformation \cite{d}
$$x_{T}(e^{2\pi i}z)=e^{-4\pi ip_{-}/Q}x_{T}(z)\;\;,\;\;\bar
x_{T}(e^{2\pi i}z)=e^{4\pi ip_{-}/Q}\bar x_{T}(z).\eqno(33)$$
The conformal weight of this twist fields is ${p_{-}\over
Q}\left(1-{2p_{-}\over Q}\right)$ \cite{d} so that the total AS
conformal weight is
$$\Delta_{II}=-2p_{+}p_{-}+{p_{-}\over Q}\left(1-{2p_{-}\over
Q}\right)\;\;,\;\;p_{-}>0\eqno(34)$$
in perfect agreement with (24b).
Type II representation operators do not carry transverse momentum.
When $2p_{-}/Q=N\in Z^{+}$ then $x_{T},\bar x_{T}$ are not twisted
but the trivial
vaccum $|0>$ has "spectraly flowed" to a new vacuum $|N>$ which
satisfies
$a_{m\geq N}|N>=0$, $\bar a_{m>-N}|N>=0$ where $a_{m},\bar a_{m}$ are
the oscillator modes of $x_{T},\bar x_{T}$.

\underline{Type III Representations}. These have a lw state at the
base and the corresponding lw operator is
$$R^{III}_{p_{+},p_{-}}=\exp\left[ip_{+}x^{-}+ip_{-}x^{+}\right]
V^{III}_{p_{-}}(x^{1},x^{2}).\eqno(35)$$
with $p_{-}<0$.
The OPE (25) implies that
$$\d x_{T} (z) V^{III}_{p_{-}}(w)\sim (z-w)^{1-{2p_{-}\over
Q}}\;\;,\eqno(36a)$$
$$\d \bar x_{T} V^{III}_{p_{-}}(w)\sim (z-w)^{-{2p_{-}\over
Q}},\eqno(36b)$$
$V^{III}$ is the conjugate twist field of $V^{II}$ and has the same
conformal
weight.
The AS conformal weight is again in agreement with (24c):
$$\Delta_{III}=-2p_{+}p_{-}-{p_{-}\over Q}\left(1+{2p_{-}\over
Q}\right)\;\;,\;\;p_{-}<0.\eqno(37)$$
Type III representations also do not carry transverse momentum.
The state multiplicities of type II and III representations for
generic $p_{-}$
are similar (up to an extra factor of $1/\eta$) to those of the
discrete
series of SL(2,R) that are not doubly degenerate, \cite{BK}.
The reason is that the null vectors in both types of representations
are the same:
they reflect the presence of a lowest (or highest) weight at
the base.
There are extra null vectors only when $p_{-}=Qm/2$ with $m\in Z$.
The determinant formula and state multiplicities for type II and III
representations will be presented elsewhere.

Some remarks are in order here.
The type II and III representations with $p_{-}=\pm Q/2$ can be
constructed without reference to twist fields.
This is also obvious from the fact that the conformal weight of the
twist field
vanishes at $p_{-}=\pm Q/2$.
In this case the full (semi-infinite) tower of base operators is
$$R^{II}_{p_{+},n}=\exp\left[i(p_{+}+n/Q)x^{-}+iQx^{+}/2\right](\d
\bar x_{T})^{n}\;\;,\;\;n=0,1,2,\cdots\eqno(38a)$$
$$R^{III}_{p_{+},n}=\exp\left[i(p_{+}-n/Q)x^{-}-iQx^{+}/2\right](\d
x_{T})^{n}\;\;,\;\;n=0,1,2,\cdots\eqno(38b)$$
The transverse space remains flat (untwisted) when $p_{-}=mQ/2$ with
$m\in Z$.
For $2p_{-}/Q$ rational it has a conical singularity, whereas for
irrational
$2p_{-}/Q$ the picture is not at all clear.

The situation above is very much similar to the string Scherk-Schwarz
mechanism, \cite{SS}.
For type II and III respresentations, we observe, that all the states
have the same
charge ($p_{+}$) under a simultaneous O(2) rotation in the tranverse
space
and a translation in $x^{-}$.
Thus they could be viewed as the untwisted sector of an orbifold of
flat space
under the action of the aforementioned transformation, like the
string Scherk-Schwarz mechanism (were the translation acts on a
compact coordinate).
The type I representations can be viewed as the twisted sector of
that orbifold.
We will show below, that the $\ec$ current algebra theory can be
obtained from
flat $M^{4}$ by a special $O(2,2)$ transformation. This type of
picture corroborates the picture we presented above.

The algebra $\ec$ has a natural notion of ``spectral flow" (or
internal automorphism) which is generated
by changing the boundary conditions of $P^{\pm}$ by an element of the
U(1)
subgroup generated by $J$:
$$P^{\pm}_{m}\to P^{\pm}_{m\pm\a}\;\;,\;\; J_{m}\to
J_{m}\;\;,\;\;T_{m}\to
T_{m}-ik\a\delta_{m,0}\eqno(39)$$
$$L^{AS}_{m}\to L^{AS}_{m}+i\a J_{m}-i\a bT_{m}+
{\a^2\over 2}Q^{2}\delta_{m,0}\eqno(40)$$
This implies the flow of conformal weights
$$L_{0}\to L_{0}-\a Q(p_{+}-p_{-})+{\a^2\over 2}Q^{2}.\eqno(41)$$
Spectral flow is useful in order to construct representations for
orbifolds of the current algebra theory.
When $\a=m\in Z$, one obtains representations of different "zero
mode"
algebras, $P^{\pm}_{\pm m}$, $T_{0}-ikm$, $J_{0}$.
These break the global symmetry under the current algebra.
They can be realized in terms of vertex operators with $p_{-}\not=0$
and $\vec p_{T}\not=0$.

What we have done so far is to map the representation theory of $\ec$
current algebra to free field theory and its orbifolds, and as such
one might wonder
if this is at all useful.
We will try to argue here that the above results are essential
in understanding string propagation in the background of a
gravitational wave
in 4-d Minkowski space.

The key remark was made in \cite{W2}, where the $\s$-model for
$E^{c}_{2}$
was constructed and its background data were interpreted as a
gravitational
wave in $M^{4}$, and an appropriate $B_{\mu\nu}$ background to cancel
the
energy of the wave.
We will use our current algebra results in order to understand the
physics
of strings propagating in such a gravitational wave background.

Following \cite{W2}, we parametrize the $E^{c}_{2}$ group element as
$g=\exp[a_{1}P_{1}+a_{2}P_{2}]\exp[uJ+vT]$.
The $E^{c}_{2}$ $\s$-model action reads \cite{W2}:
$$S_{E}={k\over 2\pi}\int \left(\d a_{i}\db a_{i}
-{1\over 2}\epsilon^{ij}a_{i}(\d a_{j}\db u +\db a_{j}\d u)+b\d u\db
u+\d u\db v+\db u\d v+u\epsilon^{ij}\d a_{i}\db
a_{j}\right).\eqno(42)$$
The measure in this coordinate system is simply $da_{1}da_{2}dudv$
which is left-right invariant.

Performing  some integrations by parts in the last term of eq. (42)
and defining $a_{1}=r\cos\theta$, $a_{2}=r\sin \theta$, we obtain the
equivalent action
$$S={k\over 2\pi}\int \left(\d r\db r +r^2\d\theta\db \theta -r^{2}\d
u\db \theta +b\d u\db u+\d u\db v+\db u\d v\right)\eqno(42')$$
If we start from flat space in the $r,\theta,u,v$ coordinates and
perform an O(3,3) transformation in the $\theta,u,v$ plane then we
obtain action (42').
The appropriate O(3,3) transformation is given by
$$\hat a=(d^{-1})^{t}=\left(\matrix{1&0&0\cr -1/2&0&b/2\cr
0&0&1\cr}\right)\;\;,\;\;\hat b=0\;\;,\;\;\hat
c=\left(\matrix{0&0&-1/2\cr
0&0&0\cr 1/2&0&0\cr}\right)$$

The model is invariant under a chiral (L+R) $E^{c}_{2}$ symmetry.
The holomorphic currents are
$$J^{1}=k(\d a_{2}-a_{1}\d u)\;\;\;,\;\;\;J^{2}=k(\d a_{1}+a_{2}\d
u)\eqno(43a)$$
$$J^{3}=k\left(\d v+{1\over 2}\epsilon^{ij}a_{i}\d
a_{j}+\left(b-{1\over 2}a_{i}a_{i}\right)
\d u\right)\;\;\;,\;\;\;J^{4}=k\d u\eqno(43b)$$
corresponding to the following symmetry transformations ($\delta
S_{E}\equiv
-{1\over\pi}\int\epsilon\db J$):
$$\delta^{1}a_{1}=0\;\;,\;\;\de^{1} a_{2}=\e\;\;,\;\;\de^{1}
u=0\;\;,\;\;
\de ^{1}v=-{a_{1}\over 2}\e,\eqno(44a)$$
$$\de^{2}a_{1}=\e\;\;,\;\;\de^{2} a_{2}=0\;\;,\;\;\de^{2}
u=0\;\;,\;\;
\de ^{2}v={a_{2}\over 2}\e,\eqno(44b)$$
$$\de^{3}a_{1}=a_{2}\e\;\;,\;\;\de^{3}a_{2}=a_{1}
\e\;\;,\;\;\de^{3}u=\e\;\;,\;\;\de^{3}v=0,\eqno(44c)$$
$$\de^{4}a_{1}=0\;\;,\;\;\de^{4}a_{2}=0\;\;,\;\;\de^{4}u=0
\;\;,\;\;\de^{4}v=\e.
\eqno(44d)$$
Similarly, the antiholomorphic currents are
$$\bar J^{1}=k(\cos u\db a_{1}+\sin u\db a_{2})\;\;\;,\;\;\;
\bar J^{2}=k(\cos u\db a_{2}-\sin u\db a_{1})\eqno(45a)$$
$$\bar J^{3}=k\left(\db v+b\db u-{1\over 2}\e^{ij}a_{i}\db
a_{j}\right)\;\;\;,\;\;\;
\bar J^{4}=k\db u\eqno(45b)$$
corresponding to the symmetry transformations
$$\dd^{1}a_{1}=\e\cos u\;\;,\;\;\dd^{1}a_{2}=\e\sin
u\;\;,\;\;\dd^{1}u=0\;\;,\;\;\dd^{1}v=(a_{1}\sin u-a_{2}\cos
u){\e\over 2},\eqno(46a)$$
$$\dd^{2}a_{1}=-\e\sin u\;\;,\;\;\dd^{2}a_{2}=\e\cos
u\;\;,\;\;\dd^{2}u=0
\;\;,\;\;\dd^{2}v=(a_{1}\cos u+a_{2}\sin u){\e\over 2},\eqno(46b)$$
$$\dd^{3}a_{1}=0\;\;,\;\;\dd^{3}a_{2}=0\;\;,\;\;
\dd^{3}u=\e\;\;,\;\;\dd^{3}v=0,
\eqno(46c)$$
$$\dd^{4}a_{1}=0\;\;,\;\;\dd^{4}a_{2}=0\;\;,\;\;
\dd^{4}u=0\;\;,\;\;\dd^{4}v=\e;
\eqno(46d)$$
$\de^{i}$, $\dd^{i}$ satisfy the $E^{c}_{2}$ algebra, as they should.

The physical interpretation of the background becomes more
transparent
in a different coordinate system \cite{W2}:
$$a_{1}=x_{1}+\cos u x_{2}\;\;,\;\; a_{2}=\sin u x_{2}\;\;,\;\;
v\to v+{1\over 2}x_{1}x_{2}\sin u\eqno(47)$$
where the action (after some integrations by parts) becomes
$$S_{E}'={k\over 2\pi}\int\left(\d x_{i}\db x_{i}+2\cos u\d x_{2}\db
x_{1}
+b\d u\db u+\d u\db v+\db u\d v\right).\eqno(48)$$
The metric in (48) is that of a gravitational wave with a null
Killing symmetry\footnote{String backgrounds with null Killing
vectors have also been considered in \cite{H}.}.
It has coordinate singularities at $\cos u=\pm 1$.
The Riemann tensor is finite, however, its only non-zero elements
being
$R_{ux_{1}ux_{1}}=R_{ux_{2}ux_{2}}=1/4$ and $R_{ux_{1}ux_{2}}=\cos u
/4$.

In order to construct a (bosonic) string vacuum, we have to tensor
this theory with a $c=22$ CFT.
The simplest case is to add 22 flat Euclidean non-compact
coordinates.
Then the background is a gravitational wave in $M^{26}$.
The spectrum contains type I representations which carry transverse
momentum
but have $p_{-}=0$. Non-trivial interactions happen only in the
tranverse space.
Type II, III representations describe particles with no trasverse
momentum but non-zero longitudinal momenta.
Here the non-trivial interactions are in the longitudinal directions.
The structure of the type I states is well understood, but some more
effort is needed in order to understand type II and III
representations and their
state multiplicities. Work in this direction is under way.

We examine below some issues of duality and cosets in the $\s$-model
picture.

The action (48) has several explicit Killing symmetries (namely
$x_{i}$ and $v$
are Killing coordinates).
It should be noted also that the $v$-independent part of the action
is that of
an SU(2) WZW model (in Euler angle parametrization) the difference
being
that here, instead of angles, we have non-compact coordinates.
The $Z_{2}$ duality on $x_{i}$ acts trivially since they are
non-compact coordinates. Duality in the $v$ coordinate (whose Killing
vector is null)
is quite peculiar and does not seem to be a symmetry.
This is not true for O(3,3,R) transformations which will produce a
(different in general) conformally invariant theory.
Up to GL(2) transformations and constant B-shifts, there is a
one-parameter
family of backgrounds that was described in \cite{hs,gk} with the
additional $u-v$ coupling, which does not change.
This family is generated by the marginal perturbation $\int J^{1}\bar
J^{1}$.

We will now describe various cosets of the $\ec$ theory.

On the Cartan, there are two possible subgroup stress tensors:
$$T_{c=2}={1\over 2k}\xx
2JT-bT^2\xx\;\;,\;\;T_{c=1}=-R\left(J-{1+2bkR\over
4kR}T\right)^2,\eqno(49)$$
where $R$ is a positive real number.
Thus, we can obtain two coset stress tensors, $T_{AS}-T_{c=2}$, with
$c=2$,
and $T_{AS}-T_{c=1}$, with $c=3$.
In the $T_{AS}-T_{c=2}$ coset, $P^{\pm}$ descend to two chiral U(1)
currents.
This implies that the theory is that of two free scalar fields or an
orbifold thereof.
In the $T_{AS}-T_{c=1}$ coset, $J+(1-2bkR)T/4kR$ survives as a chiral
U(1) current whereas $P^{\pm}$ become parafermionic with conformal
weight $R$.

There is another pair of stress tensors. The first we obtain only on
the $P_{1,2}$ part of the algebra:
$T_{P}=\left(\cos\gamma P_{2}+\sin\gamma P_{1}\right)^2/2k$,
where $\gamma$ is a free parameter and $c=1$.
Again $T_{AS}-T_{P}$ is another coset, with $c=3$.

Of course the existence of these cosets is trivial from the point of
view of the free
field representation of the current algebra (however, the solution of
the last coset is not that trivial).
Their $\s$-model realizations are much less trivial, as we shall see.

We will describe here the $\s$-model picture of the last coset (with
$c=3$).
This will be obtained by gauging the symmetry $x_{1}\to x_{1}+\e$,
$x_{2}\to x_{2}+\e$ in the action (48).

The gauged WZW model action is
$$S(A)=S_{E}'+{k\over 2\pi}\int\left[A(\db x_{2}+\cos u \db x_{1})+
\bar A(\d x_{1}+\cos u\d x_{2})+{1\over 2}A\bar A(1+\cos
u)\right].\eqno(50)$$
By integrating out the gauge fields, and gauge fixing
$x_{1}+x_{2}=0$,
we obtain the 3-d $\s$-model action
$$S_{3}={k\over 2\pi}\int \left[{1-\cos u\over 1+\cos u}\d x\db x+b\d
u\db u+
\d u\db v+\db u\d v\right]\eqno(51)$$
and the dilaton $\Phi =-{1\over 2}\log(1+\cos u)$ coming from the
quadratic term of the gauge fields \cite{k4}.
It is not difficult to check that, for this background, the 1-loop
$\beta$-
function equations and $\delta c$ vanish.

It should be remarked here that, as in the case of the 4-d action,
the action (51) is that of the SU(2)/U(1) coset (up to the
non-compactness
of the Euler angles) coupled to $v$.

The metric in (51) describes a gravitational wave in (2+1)
dimensions,
the only difference from the 4-d case being that we now have
curvature singularities.
The only non-zero component of the Riemann tensor is
$R_{xuxu}=-(1-\cos u)/
(1+\cos u)^2$; it has a curvature singularity at $\cos u =-1$.
This is a naked singularity.

To conclude, we have given the representation of the $\ec$ current
algebra
and provides an important step towards the solution of the
$E^{c}_{2}$ $\s$-model
by mapping it to free fields. An interesting structure appears
similar to the $SL(2,R)$ representation theory.
In the associated string theory, there are two kinds of particles.
Those that have momentum in the transverse directions, where their
interactions take place and those that have only longitudinal
momentum.
It seems of greater interest to understand the interactions and the
full spectrum of the theory in more detail.
The supersymmetric generalization should also be considered in order
to obtain
a tachyon-free spectrum.

\vskip 2cm
\noindent
{\bf Acknowledgments}

We would like to thank I. Bakas for many illuminating
discussions.
Part of this work was supported by EEC grants SC1$^{*}$-0394C and
SC1$^{*}$-CT92-0789.\\

\end{document}